\documentclass{article}
\usepackage{amsmath}
\usepackage{graphicx}
\usepackage{natbib}
\usepackage{url} 
\usepackage{graphicx}
\usepackage{amsmath}
\usepackage{natbib}
\usepackage{amscd,amssymb,amsfonts,verbatim}
\usepackage{mathrsfs}
\usepackage{bbm}
\usepackage{listings}
\usepackage{epsfig}
\usepackage{enumitem}
\usepackage{url}
\usepackage{booktabs,makecell,ltablex}
\usepackage{multirow}
\usepackage{algorithm,algpseudocode}
\usepackage{array}

\usepackage{tikz}
\usetikzlibrary{patterns}
\usetikzlibrary{trees}
\usepackage{pgfplots}
\usetikzlibrary{svg.path}
\usepgfplotslibrary{fillbetween}
\usetikzlibrary{shapes.misc}
\usetikzlibrary{shapes.geometric}
\usetikzlibrary{arrows}
\pgfplotsset{compat=1.16}
\pgfplotsset{soldot/.style={color=blue,only marks,mark=*}}
\pgfplotsset{holdot/.style={color=blue,fill=white,only marks,mark=*}}
\newcommand\BibTeX{{\rmfamily B\kern-.05em \textsc{i\kern-.025em b}\kern-.08em
		T\kern-.1667em\lower.7ex\hbox{E}\kern-.125emX}}

\def\E{\mathbb{E}}

\newcommand{\indep}{\perp \!\!\! \perp}

\graphicspath{{Figs/}}

\makeatother

\begin{document}

\def\spacingset#1{\renewcommand{\baselinestretch}%
{#1}\small\normalsize} \spacingset{1}


  \title{\bf 
A scalable Bayesian double machine learning framework, with application to racial disproportionality assessment}
  \author{Yu Luo\thanks{
  		Department of Mathematics, King's College London, United Kingdom} ,  \hspace{.2cm}
  		Vanessa McNealis \thanks{
  			School of Mathematics and Statistics, University of Glasgow, United Kingdom} ,  \hspace{.2cm}
  	Yijing Li\thanks{
  		Department of Informatics, King's College London, United Kingdom} \hspace{.2cm}
  }
 \date{ }
  \maketitle

\bigskip
\begin{abstract}
Racial disproportionality in stop and search practices elicits substantial concerns about its societal and behavioral impacts. In London, Black individuals are about four times more likely to be stopped and searched than White individuals. Using data on stop and search events in London from January 2019 to December 2023, this paper aims to investigate disproportionality in the volume of stops for expressive crimes involving Black individuals compared to other ethnicities. We employ a semi-parametric partially linear structural regression method and introduce a Bayesian empirical likelihood procedure combined with double machine learning techniques  to control for high-dimensional confounding and to accommodate strong prior assumptions.  In addition, we show that the proposed procedure yields a valid posterior in terms of coverage. Applying this approach to the stop and search dataset, we find that racial disproportionality aimed at the Black community may be influenced by the borough racial composition when focusing on expressive crimes. 
\end{abstract}

\noindent%
\vfill

\newpage
\spacingset{1.5}

\section{Introduction}
\subsection{Background}
Racial disproportionality in stop and search practices, particularly in cities such as London, raises important questions about its broader social and behavioral effects. Expressive crimes, such as vandalism, public disorder, or gang-related violence, are often argued to be influenced by social alienation, perceived injustice, or strained community-police relations, which can result from disproportionate policing practices.

\cite{bowling2007disproportionate} suggested that racial discrimination within policing is exacerbated by the stop and search police powers in the UK targeting Black individuals disproportionately. The UK's State of Policing report in 2022 highlighted that stop and search is an important tool for preventing and detecting crime, with significant public support when used fairly and proportionately, particularly in targeting weapons and drugs \citep{hminspec2022}. \cite{Tiratelli2018ssdisproportionate} assessed the effectiveness of stop and search in reducing crime in London, with findings indicating some correlation between stop and search and drug offenses, but relatively limited impact on broader crime reduction.

In recent years, the issue of disproportionality in stop and search practices has remained a significant concern in London, where the disparities between ethnic groups have persisted despite various reform efforts. The March 2024 Disproportionality Board Data Pack \citep{mopac2023}, part of the Mayor’s Action Plan for Transparency, Accountability, and Trust in Policing, highlighted that Black individuals in London are still 3.3 times more likely to be stopped and searched compared to White individuals, and this figure rises to 6.2 times for searches related to weapons. 

Scholars have identified that stop and search events are highly concentrated in low-income neighborhoods or areas with higher populations of minority ethnic groups. According to \cite{miller2020}, the increased use of new surveillance technologies, such as predictive policing software, has transformed policing in cities such as London, intensifying the monitoring of particular demographic groups. \cite{suss2023} and \cite{meng2017} demonstrated that the spatial patterns of stop and search in London are closely linked to areas characterized by higher levels of economic inequality and minority populations, revealing racial bias embedded in police practices. \cite{ober2022} argued that in France, Germany and other European cities, police actions against adolescents in certain neighborhoods are often shaped by institutional biases tied to economic deprivation and race. Given these findings, it is crucial to use a data-driven approach to quantify the effect of racial disproportionality in stop and search. For example, a robust statistical approach is needed to account for the role of various confounding factors, such as differences in demographics across communities, socioeconomic conditions and policing priorities, from potential biases in police decision-making.

\subsection{Motivating example: Racial disproportionality assessment in Stop and Search}
\label{data}
In this paper, we investigate a the stop and search dataset comprising individual stop and search monthly records in London from January 2019 to December 2023, including date and time, street-level location, ethnicity, gender and age of the person stopped, legislation, object searched and outcome.  Further description of this dataset can be found in the Appendix \ref{appn}.  Based on 2.6 million stop and search incidents aggregated within the borough level, the objective of this study is to examine how the level of disproportionality in stop and search practices for expressive crime targeting Black individuals varies across London boroughs. That is, the focus is on borough-level inequality in policing outcomes rather than on the overall presence or absence of racial bias across the city. Even if disproportionality is high across London, the analysis seeks to identify whether certain boroughs exhibit relatively greater or lesser disproportionality than the average level in London.

Following \cite{bowling2007disproportionate}, disproportionality is defined as the ratio between an observed rate and the expected rate based on the population at risk. In this context, the expected rate serves as a benchmark that reflects what would be observed under proportionality with respect to population size. Therefore, we define the disproportionality index (DI) for expressive crime targeting Black individuals in each borough $i$ (total 33 Boroughs) as: 
	\[
	\text{DI}_i = \dfrac{\overline{S}^{\,B}_{i} / N^{\,B}_{i}}{\overline{S}_{\text{London}} / N_{\text{London}}},
	\]
	where $\overline{S}^{\,B}_{i}$ represents annual average number of stop and search incidents involving Black individuals for expressive crime in borough $i$, and $N^{\,B}_{i}$ represents Black resident population in borough $i$, and $\overline{S}_{\text{London}}$  and $N_{\text{London}}$ represent the annual average count and total population across London respectively, aggregated over all racial groups. The denominator represents the overall stop and search rate across the entire London population, irrespective of race. We believe this definition provides an expected rate under a benchmark in which stop and search is proportional to population size and does not vary systematically across demographic groups or boroughs. Under this interpretation, the denominator reflects the rate that would be expected for any subgroup, including Black individuals, if policing powers were applied uniformly across the population. The index DI$_i$ captures the extent to which the observed rate for Black individuals deviates from this population-wide rate. $\text{DI}=1$ indicates no disproportionality, whereas $\text{DI}<1$ indicates a stop and search rate below expectation while $\text{DI}>1$ indicates a stop and search rate above expectation, suggesting that Black individuals are subject to stop and search at a rate above the population-wide expected benchmark.

\begin{figure}[ht]
	\centering
	\includegraphics[width=0.6\textwidth]{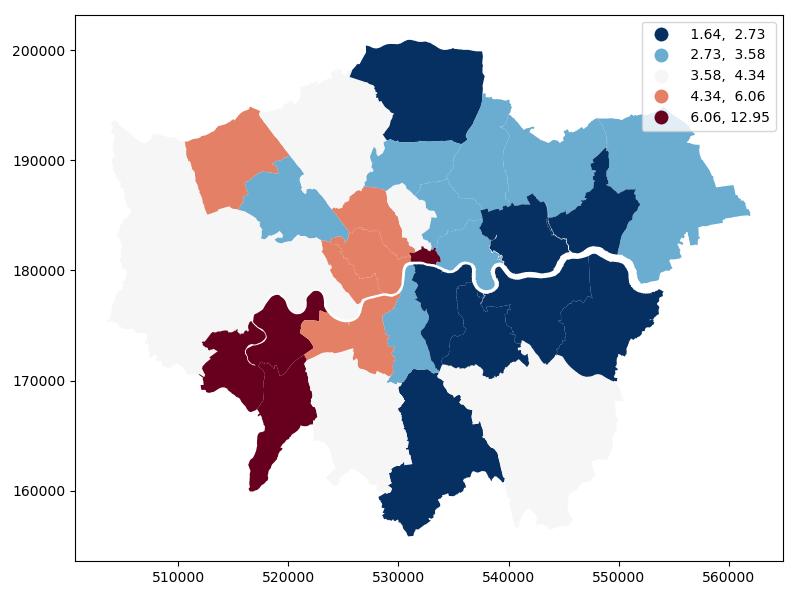}
	\caption{ \label{DI} DI for Black People in Stop \& Search for Expressive Reasons}
\end{figure}

Figure \ref{DI} illustrates the DI across  boroughs in London. The South-West London boroughs of Richmond upon Thames and Kingston upon Thames and the City of London have the largest DI values. These boroughs have considerably larger index values, revealing stop and search rates above expectation among the Black population. In contrast, east end London boroughs, especially the riverside boroughs — Lewisham, Greenwich, Bexley, Newham, and Barking and Dagenham — all have low DI scores. Boroughs nearer the perimeter of the city, like Enfield and Croydon, also have lower DI scores. This trend indicates a gradient in the distribution, with more central and wealthy areas tending to have larger index values and outer or more residential boroughs having comparatively lower scores.

Having defined the DI as the outcome of interest, we use the percentage of Black people in each borough as the treatment variable. This allows to investigate the effect of borough racial composition on disproportionality in stop and search. Specifically, areas with a higher percentage of Black residents might experience more police attention, which would inflate the DI. If policing practices were biased or influenced by racial profiling, the stop and search rates for Black individuals might rise disproportionately, driving up the DI. Moreover, in boroughs with larger Black populations, more recorded crimes against Black individuals could naturally occur, but if the rate of police actions exceeds the expected level in London based on population share, it would signal disproportionality. For example, if Black residents make up 30\% of a borough but account for 60\% of stop and search incidents, the DI would be elevated, demonstrating a disproportionate impact.

The proposed estimand, the effect of the share of Black residents on a borough-level DI, differs from some of the existing literature, which often examines the effect of an individual’s (perceived) race on police actions (e.g., use of force). For instance, the post-treatment variable considered in \cite{zhao2022note}, police detainment, can be viewed as an individual-level analogue of the outcome considered in this paper, since the DI is constructed from the frequency of ``detainments" or stop interactions involving individuals from minority backgrounds. One key distinction is that our analysis is conducted at the areal level, whereas \cite{knox2020administrative,zhao2022note} focus on individual-level observations. Moreover, our estimand does not directly contrast outcomes for minority and non-minority individuals. Instead, it asks how the level of disproportionality affecting Black individuals varies across boroughs with different racial compositions. In this sense, the estimand shifts the focus from individual-level differential racial treatment to
	geographic variation in disproportionality across boroughs. From a policy perspective, the investigation of borough racial composition on geographical disproportionality can help identify boroughs at risk and structural mechanisms requiring more attention from authorities, including police deployment practices and stop and search patterns. However, they should not by themselves be interpreted as definitive evidence of individual-level discrimination.

\subsection{Statistical challenges and related work}
Disparities in institutional decision-making have significant implications for both individual well-being and broader societal outcomes. In healthcare, unequal access to treatments or disproportionality in clinical decision-making across population groups can lead to persistent health inequities. Similarly, differential treatment by law enforcement in policing may raise important questions about its broader social and behavioral effects. Quantifying racial bias in policing has become a central question in policy-oriented social science and applied statistics. Despite wide availability of administrative policing data, such as stop records, arrest databases, and officer-reported activity logs, using these data to make credible causal inferences about discrimination remains methodologically challenging. 

From a causal inference perspective, investigation of borough composition on racial disproportionality is first complicated by confounding of the exposure-outcome association. Confounding exists whenever exposure (or treatment) assignment is dependent on predictors that also influence the outcome, and appropriate adjustment is required to estimate the corresponding effect. Variables, such as the number of schools and proportion of green space, can obscure the true relationship between these variables.  For example, variation in areal stop and search rates involving Black individuals could also influenced by other factors than the exposure of interest, such as the proportion of educated households and the number of schools in the area. Determining this effect in this complex scenario requires careful consideration of variables and the implementation of rigorous research in data analysis. Recent methodological advances seek to formalize these adjustments. \cite{gaebler2022causal} argue that many policing studies lack explicit causal models, leading to implicit identifying assumptions. \cite{jung2018mitigating} propose a risk-adjusted regression framework to mitigate included- and omitted-variable bias and to focus disparity estimates on differences not explained by legitimate police-relevant risk factors. Recently, \cite{huang2024causal} introduce a mobility-adjusted causal estimand, demonstrating that individuals’ movement patterns fundamentally shape exposure to policing. 

Another potential threat to identification in this context is selection bias, since stop and search events often arise because of a form of racial bias. The data we get to observe arise from interactions the police have with the civilians they choose to stop, and are certainly not a random sample of all police-civilian interactions. \cite{knox2020administrative} demonstrate that administrative policing datasets inherently condition on the outcome of police discretion, and show that estimands targeting racial bias that condition on being stopped can potentially have the opposite signs as the estimands of interest. A policing agency could appear unbiased or even favorable to minority groups among stopped individuals while simultaneously discriminating in its choice of whom to stop. Based on this, \cite{zhao2022note} clarify the definition of discrimination in causal inference terms and emphasize the need for estimands that avoid conditioning on post-treatment variables. Their work stresses that identifying racial disparities requires not only conditioning strategies but careful articulation of what causal question is being answered. As mentioned in Section \ref{data}, the causal estimand in this work avoids the specific form of post-treatment conditioning considered by \cite{knox2020administrative} and \cite{zhao2022note}, as we do not compare subsequent outcomes among individuals who have been stopped. Instead, the stop and search process contributes directly to the borough-level outcome, through the DI. That being said, the observed events are still shaped by police decisions about whom and where to stop. Therefore, the DI considered here should be interpreted as an areal-level measure of observed geographic disproportionality in stop and search activity, rather than as conclusive evidence of individual-level discrimination.

To estimate causal effects in complex, high-dimensional settings, applied researchers often adopt flexible semiparametric models. A commonly used tool is the partially linear regression framework \citep{robinson1988root}, which specifies separate regression functions for the treatment and outcome. The model for the treatment variable is often termed as `propensity score' model. Propensity score adjustments \citep{rosenbaum1983central} have been extensively used to reduce confounding bias in estimating causal effects.  The propensity score is defined as the conditional probability of receiving treatment given confounding covariates. The propensity score can be used to break the dependence between confounders and exposure, to create balance in the distribution of confounders across exposure groups, and to allow valid causal inference. However, a key statistical challenge in applying partially linear regression for stop and search practices is the presence of high-dimensional confounders, where the parameter space for nuisance parameters grows with the sample size. In this case, traditional semi-parametric theory might not offer valid inference for the parameter of interest. To overcome these challenges, \cite{chernozhukov2018double} proposed the use of machine learning (ML) methods to estimate the nuisance parameters and provided a simple and root-$n$ consistent procedure to estimate the parameter of interest via the Neyman moment equation and sample splitting.

Another challenge arises, when analyzing stop and search data, due to the strong prior beliefs and assumptions often held by policymakers and law enforcement agencies. These priors may encode operational experience or historical evidence, but they may also reflect entrenched practices or anecdotal reasoning. Ignoring such prior information risks disconnecting statistical analysis from the policy context in which decisions are made; however, incorporating it uncritically can reinforce existing biases. This consideration motivates a framework that can both formally incorporate and transparently interrogate prior beliefs. Therefore in this case, a Bayesian approach offers advantages relative to standard frequentist methods. First, it provides a mechanism for incorporating policy-relevant prior information into the analysis. This is particularly useful in policing applications where historical data, expert elicitation, or institutional knowledge play a central role in shaping expectations. Second, unlike approaches that rely on large-sample approximations, Bayesian inference provides coherent uncertainty quantification even in small-sample settings, which commonly arise in area-level analyses. Third, Bayesian approach provides direct posterior probability summaries of scientifically relevant causal contrasts, such as differences in the posterior probability of being stopped between Black and White individuals in a given borough, after adjusting for confounders. These considerations motivate the development of a Bayesian framework to our application. Specifically,  we aim to incorporate Bayesian inference within a double ML framework to obtain coherent probabilistic statements about the effects of disproportionality, enhance the stability of estimation, and allow the incorporation of domain knowledge. This offers a coherent updating framework for regularization and prior information, which are valuable in high-dimensional or weak-signal settings common in health and social policy applications. There is growing interest in the application of Bayesian methodology to two-stage propensity score regression analysis;  however, most of existing Bayesian approaches require a full parametric specification for both structural equations in both regression models \citep[see, for example][]{mccandless2010cutting,kaplan2012two}.  There are   {several} works already attempting to solve this problem via a semi-parametric perspective \citep[for example,][]{graham2016approximate,liu2020estimation}; these methods typically exploit the Bayesian bootstrap \citep{rubin1981bayesian,chamberlain2003nonparametric} to perform inference. Recently,  \cite{luo2021bayesian} proposed to draw inference for the posterior from a Bayesian predictive distribution via a Dirichlet process model, extending the Bayesian bootstrap, and opening up the possibility of performing doubly robust causal inference based on a non-parametric specification. In their work, the posterior samples are generated by resampling weights from the P\'{o}lya urn. It links with recent advancements in Bayesian empirical titled likelihood \citep{chib2018bayesian,yiu2020inference,luoemp2021}, where the weights are replaced by the empirical probabilities. \cite{antonelli2022causal} propose to use the Gaussian process (GP) regression, and then the MCMC estimate is plugged into estimating equation. Recently, \cite{ditraglia2025bayesian} introduce a fully parametric Bayesian approach for causal inference in partially linear models with high-dimensional control variables. However, despite these advances, computational efficiency and theoretical guarantees remain open challenges, where covariates may be high dimensional and the nuisance components may be complex. In this paper, we aim to marry the Bayesian method and the semi-parametric double ML framework and provide a computationally efficient procedure to draw inference on the parameter of interest. We argue that our method bridges this gap by demonstrating how the approximate frequentist distribution theory can find an equally effective interpretation within a Bayesian framework in the high-dimensional setting. In addition, we show that the proposed procedure generates a valid posterior according to \cite{monahan1992proper}, indicating a valid putative ‘posterior’ density computed by a non-standard method should still make probability statements consistent with Bayes’ rule.

The remainder of this paper is organized as follows. In Section \ref{sec:stat}, we articulate the causal question for the study and define the associated causal estimand. In addition, we review the estimation procedure of partially linear regression, and introduce the notion of approximating Bayesian formulations and how to perform inference via the Neyman moment equation.  We verify the validity of the proposed posterior inference in the spirit of \cite{monahan1992proper} in Section \ref{sec:valid}.   Section \ref{sec:sim} shows some simulation examples which compare the proposed method with some other frequentist and Bayesian approaches, following with  stop and search data analysis in Section \ref{sec:app}.  Finally,  Section \ref{sec:dis} presents some concluding remarks and future research directions.

\section{Methodology}
\label{sec:stat}

\subsection{Notation and causal assumptions}

Let $\left\{Z_i = (Y_i,D_i,X_i),i = 1,\ldots,n\right\}$ denote independent and identically distributed data measured on $n$ study units (e.g., different boroughs), where $Y_i$ denotes an observed outcome (e.g., the disproportionality index DI), $D_i$ represents an exposure or treatment, potentially continuous (e.g., percentage of Black people in an area) with support $\mathcal{D}$, and $X_i$ is a vector of potential confounders measured on unit $i$. The primary focus of our inference is on average potential outcomes (APOs) for a population under a hypothetical exposure. We
	adopt the potential outcome framework with continuous exposures introduced by \cite{hirano2004propensity}, and let $Y_i(d)$ be the (potential) DI that would be observed in borough $i$ if the proportion of the Black population were set to $d$. Additionally, we assume differentiability and smoothness of the potential outcomes $Y_i(d)$, that is, we assume that for each $i$, the potential outcome $Y_i(d)$ is differentiable with respect to $d$ and its derivative $Y_i'(d) := \frac{\partial}{\partial d} Y_i(d)$ is continuous and bounded for all $d \in \mathcal{D}$ \citep{bong2023local}. We set the target estimand as the population average causal derivative effect (ACDE), i.e., $ \delta(d_0)= \E \left[\frac{\partial Y(d)}{\partial d}\right]_{d = d_0} = \mathbb{E}[Y'(d_0)]$ capturing the local effect of nearby treatment $d_0$ on the outcome, where the expectation is taken with respect to the data generating process of $Z$. In the context of our empirical analysis, the datum is any London borough $i$, $i = 1, \ldots, n$, so that this expectation is taken over the borough distribution, and the estimand $\delta(d_0)$ captures the local effect of borough-level racial composition on the degree of disproportionality in stop and search incidents involving Black individuals for expressive crimes. Given confounders $X_i$ in borough $i$, identification relies on the usual core causal assumptions: consistency, Stable Unit Treatment Value (SUTVA), no unmeasured confounding (NUC), and strict positivity \citep{rosenbaum1983central, hirano2004propensity}.
Under these assumptions, the distribution of potential outcomes is identified from observed data via the g-formula: $E \left[\delta(d_0)\right] = \E_X \left\{\E \left[  \delta(d_0)| D=d_0, X\right] \right\}$. In the next section, we will specifically focus on partially linear regression to estimate this causal effect using a Bayesian double ML approach.

\subsection{Partially linear regression}
To estimate the effect of policy, i.e. racial dispropotionality, we consider the partially linear regression described in \cite{robinson1988root}: 
\begin{equation}
	\label{plr}
	Y = \mu(X) + \beta D + U, \;\;\; \E(U|X,D) = 0,\;\;\; \E(D|X) = \pi (X)\\
\end{equation}

where $Y$ is the outcome variable, $D$ is the treatment variable, $X = \left(X_1,\ldots,X_p\right)$ is a list of counfounders and $U$ represents the unobserved error term. 

Under the partially linear regression model, the ACDE corresponds to the parameter $\beta$ for all $d_0 \in \mathcal{D}$. Thus, we shall refer to $\beta$ as the primary inferential target from this point onwards. 
In frequentist inference, it is a well known fact that the estimator of  $\beta$ is consistent if $\mathbb{E}\left[Y\left|X,D\right.\right]$ is correctly specified.  Additionally, a doubly robust estimator can be constructed by specifying both the outcome mean and propensity score models and combining them for estimation. To implement such inference, we need to specify the following two models: 
\begin{itemize}
	\item Propensity score (PS) model: $\E(D|X)$, which represents the conditional distribution of treatment given the confounders;
	\item Outcome regression model: $\mathbb{E}\left(Y\left|X,D\right.\right)$, which represents the conditional distribution of the outcome given the treatment variable and confounder under the observational study.
\end{itemize}
As noted in \cite{lee2018simple}, The model in \eqref{plr} can be rewritten as a new regression model:  
\begin{equation}
	\label{plr1}
	Y - \E[Y|\pi(X)] = \beta [D- \pi(X)] + V, \;\;\; V = U+ \mu(X) - \E[\mu(X) | \pi(X)] = 0,\;\;\; V \indep D |\pi(X)\\
\end{equation}
and  therefore $\E(V|\pi(X)) = 0$. This implies that $cov(D- \pi(X), V) = 0$, and the treatment effect parameter, $\beta$, can be estimated via the least square estimation procedure from regressing $Y - \E[Y|\pi(X)]$ on $D- \pi(X)$. However, \cite{hahn1998role} demonstrated that the estimator is not semiparametrically efficient when both $\E[Y|\pi(X)]$ and $\pi(X)$ are estimated non-parametrically. To address this, \cite{chernozhukov2018double} proposed estimating $\beta$ using the Neyman-orthogonal moment equation, which achieves the semi-parametric efficiency bound as it protects the estimator against first-order bias from nuisance function estimation errors. That is, $\beta$ is the solution $ \E [\psi(Z;\beta)]=0$, where 
$$\psi(Z;\beta)= [D - \pi(X)] [Y - \beta D - \mu(X)].$$
We can estimate nuisance functions, $\pi(X)$ and $\mu(X)$, non-parametrically or using ML methods regardless of the dimension of $X$.  In this way, we can reduce the high-dimensional problem to a single parameter problem if we assume $D$ is a scalar.  This approach requires that we have access to both $\mu(X)$ and $\pi(X)$. The challenge arises in the high-dimensional setting, i.e., $p >n$; however, both of these functions can be estimated via ML methods. This double ML specification gives  us  flexibility to specify the structures of $\mu(X)$ and $\pi(X)$ while focusing on the treatment effect.  However, it would be difficult to perform conventional Bayesian analysis with an strong prior input from the policy makers as there is no distribution assumption in this semi-parametric procedure. Conventional Bayesian inference focuses on updating prior belief in light of the data, and the data are summarized in the form of the likelihood. The relationship between prior beliefs and observable random quantities, $\mathbf{z}=\left(z_1,\ldots,z_n\right)$, is formulated via the de Finetti representation, i.e.,
\[
f\left(z_1,\ldots,z_n\right) = \int \prod_{i=1}^{n} f(z_i | \beta ) \pi_0(\beta) d \beta.
\] 
In the de Finetti representation, a full probabilistic model, $f(z_i | \beta )$,  is required, and $\pi_0(\beta)$ is the prior belief about $\beta$.  In this case, we have to examine procedures for Bayesian inference in the case where we wish to perform analysis of an approximate model, acknowledged to be misspecified compared to the data generating model. In the next section, we will seek solutions to incorporate the Neyman-orthogonal approach into a fully Bayesian procedure and demonstrate how the approximate frequentist distribution theory, which rests on moment constraint assumptions about distributions, can find an equally effective interpretation within a Bayesian framework.

\subsection{Bayesian non-parametric procedure for the double machine learning method}
\label{sec:BEL}
Suppose we assume that there is a set of Neyman-orthogonal equations $\psi(Z;\beta)$ such that $\E [\psi(Z;\beta)]=0$, for all $\beta$. The objective is to find a non-parametric approximate likelihood, $p(z)$, to the true data generating model $f(z|\beta)$. Therefore, we can define some discrepancy $\delta (f|p)$, subject to $\int\psi(z;\beta)  p(z)dz=0$ and $\int p(z)dz=1$.  This becomes an optimization  problem:
\begin{equation}
	\label{app}
	\min_{p}  \delta (f|p)
	\text{ subject to } \int\psi(z;\beta)  p(z)dz=0\;\; \int p(z)dz=1,\;\; \forall \beta \in \mathbb{R}.
\end{equation}
If we specify $p$ as nonparametric, and use $\delta (f|p_1,\ldots,p_n)$ to measure the discrepancy between the true data generating model and the approximating model. Specifically, $p_i$ can be the solution of the dual formulation which satisfies certain moment conditions and can be summarized as  the following constrained optimization problem 
\begin{equation}
	\label{gel}
	\min_{p_1,\ldots,p_n} \delta (f|p_1,\ldots,p_n)
	\text{ subject to } \sum_{i=1}^{n}p_i=1,\;\; p_i\ge 0,\;\; \sum_{i=1}^{n}p_i\psi(z_i;\beta)= 0.
\end{equation}
In light of \cite{read2012goodness}, the the empirical Cressie-Read statistic can be used as a goodness-of-fit  measure for  discrete multivariate data. Then we specify $\delta (f|p_1,\ldots,p_n)\equiv\text{CR}(p)$, where 
\[
\text{CR}(p)=  \frac{2}{\lambda(1+\lambda)} \sum_{i=1}^{n}\left[(np_i)^{-\lambda}-1\right] , \;\;\; -\infty<\lambda<\infty
\]
where $\lambda$  is a user-specified parameter.  From \cite{baggerly1998empirical}, this function can be rewritten as 

\[
\text{CR}(p) = \left\{\begin{array}{cc}
	-2 \sum_{i=1}^{n} \log (np_i), & \lambda= 0\\
	2n \sum_{i=1}^{n} p_i\log (np_i), & \lambda= -1\\
	\frac{2}{\lambda(1+\lambda)} \sum_{i=1}^{n}\left[(np_i)^{-\lambda}-1\right],& \lambda \ne -1 \text{ or } 0.\\  
\end{array}
\right. 
\]
Therefore, the solution in \eqref{gel} is the nonparametric likelihood which seeks to reweight the sample so that it can also satisfy the moment condition \citep{qin1994empirical}. It has been proven to possess many properties of the conventional parametric  likelihood theory \citep{owen2001empirical}. A class of generalized empirical likelihood functions are studied in \cite{imbens1998information,chernozhukov2003mcmc,newey2004higher}.   In our example, we want to place a prior on the treatment effect parameter, $\beta$, directly, and update this prior in light of the data observed. Therefore, we can replace $f\left(z_i\left|\beta\right.\right)$ with the profile likelihood, $p_i$, and then the posterior distribution for $\beta$ becomes 
\[
\pi\left(\beta\left|\mathbf{z}\right.\right)\propto\pi_0\left(\beta\right)\prod\limits_{i=1}^{n}  p_i.
\]
In this way, we can incorporate a fully Bayesian procedure for $\beta$ while satisfying the conditions in \eqref{gel}.  There are several choices of $\lambda$ leading to the existing empirical likelihood methods. For example, The case $\lambda= 0$ yields the  empirical likelihood (EL) case, where $p_i$ is obtained through the maximum likelihood estimation.  When $\lambda=-1$, it yields the exponentially tilted empirical likelihood (ETEL) minimize, where  $(p_1 , \ldots , p_n)$ minimizes the Kullback-Leibler (KL) divergence between $(p_1 , \ldots , p_n)$ and the empirical probabilities $(1/n,\ldots,1/n)$. This case has been extensively studied  under model misspecification \citep{,chib2018bayesian,yiu2020inference,luoemp2021}.  In addition, if $\lambda=-1/2$, it gives  for the Hellinger distance (HD) measure discussed in  \cite{kitamura2013robustness}. As noted by \cite{baggerly1998empirical}, a unique solution exists for \eqref{gel}, provided that zero is inside the convex hull of $\psi(z_i;\beta)$ for a given $\beta$. Applying the  Lagrange multiplier approach, we can obtain the solution to  \eqref{gel} by minimizing
\[
\text{GEL}(p)= \frac{2}{\lambda(1+\lambda)} \sum_{i=1}^{n}\left[(np_i)^{-\lambda}-1\right]+ \kappa_1\left(\sum_{i=1}^{n}p_i-1\right)+n\kappa_2^\top \sum_{i=1}^{n}p_i\times\psi(z_i;\beta)
\]
where $\kappa_1$ and $\kappa_2$ are the Lagrange multipliers.  Setting $\partial \text{GEL}/\partial p_i =0$  yields the extreme of the form 
\begin{equation}
	\label{gelsol}
	p_i= \left\{\begin{array}{cc}
		\frac{1}{n}\left[1+s + t\psi(z_i;\beta) \right]^{-1/(1+\lambda)}, & \lambda \ne -1\\
		s \exp\left[t\psi(z_i;\beta)\right], & \lambda= -1\\
	\end{array}
	\right.
\end{equation}
where $s$ and $t$ are normalized constant and determined by 
\[
\frac{1}{n} \sum_{i=1}^{n}\left(1+s+t\psi(z_i;\beta) \right)^{-1/(1+\lambda)}=1, \;
\frac{1}{n} \sum_{i=1}^{n}\left(1+s+t\psi(z_i;\beta) \right)^{-1/(1+\lambda)} \psi(z_i;\beta)=0. 
\]
Therefore, by  substituting \eqref{gelsol} into the posterior distribution, we obtain the following posterior distribution
\begin{equation}
	\label{postgel}
	\pi(\beta \left|\mathbf{z}\right.)\propto \pi_0(\beta) \times \left\{\begin{array}{cc}
		\prod_{i=1}^{n}	\frac{1}{n}\left[1+\hat s(z,\beta) + \hat t(z,\beta)\psi(z_i;\beta)\right]^{-1/(1+\lambda)} & \lambda\ne -1\\
		\prod_{i=1}^{n}\frac{\exp\left(\hat t(z,\beta)^{\top} \psi(z_i;\beta)\right)}{\sum_{j=1}^{n}\exp\left(\hat t(z,\beta) \psi(z_j;\beta)\right)} & \lambda= -1\\
	\end{array}.
	\right.	
\end{equation}
The following algorithm summarizes the computation step to the Bayesian generalized empirical likelihood method.
\begin{algorithm}[H]
	\caption{\label{A1} Algorithm to obtain the posterior sample of $\beta$ via the Bayesian generalized empirical likelihood.}
	\begin{algorithmic}[1]
		\Require{$\mathcal{D}=(z_1,\ldots,z_n)$}
		\State Estimate $\pi(x)$ and $\mu(x)$ using some ML methods.
		\For {$j$ \textbf{to} $1:J$}
		\State	Sample $\beta^{(j)} \sim \pi_j (\beta \left|\mathbf{z}\right.)$ using the MCMC approach, 
		where $\pi_j(\beta\left|\mathbf{z}\right.)\propto \pi_0(\beta) \times  \prod_{i=1}^{n}p_i^{(j)}$.
		\State $\left(p^{(j)}_1,\ldots,p^{(j)}_n\right)$ is the solution to
		$
		\min_{p_1,\ldots,p_n}  \text{CR}(p)
		\text{ subject to } \sum_{i=1}^{n}p_i=1,\;\; p_i\ge 0,\;\; \sum_{i=1}^{n}p_i\psi(z_i;\beta^{(j-1)})= 0
		$
		\EndFor
		\State \Return $\left(\beta^{(1)},\ldots,\beta^{(J)}\right)$.
	\end{algorithmic}
\end{algorithm}

\subsection{The role of sample splitting procedures}
\begin{figure}
	\centering
	\includegraphics[scale=0.75]{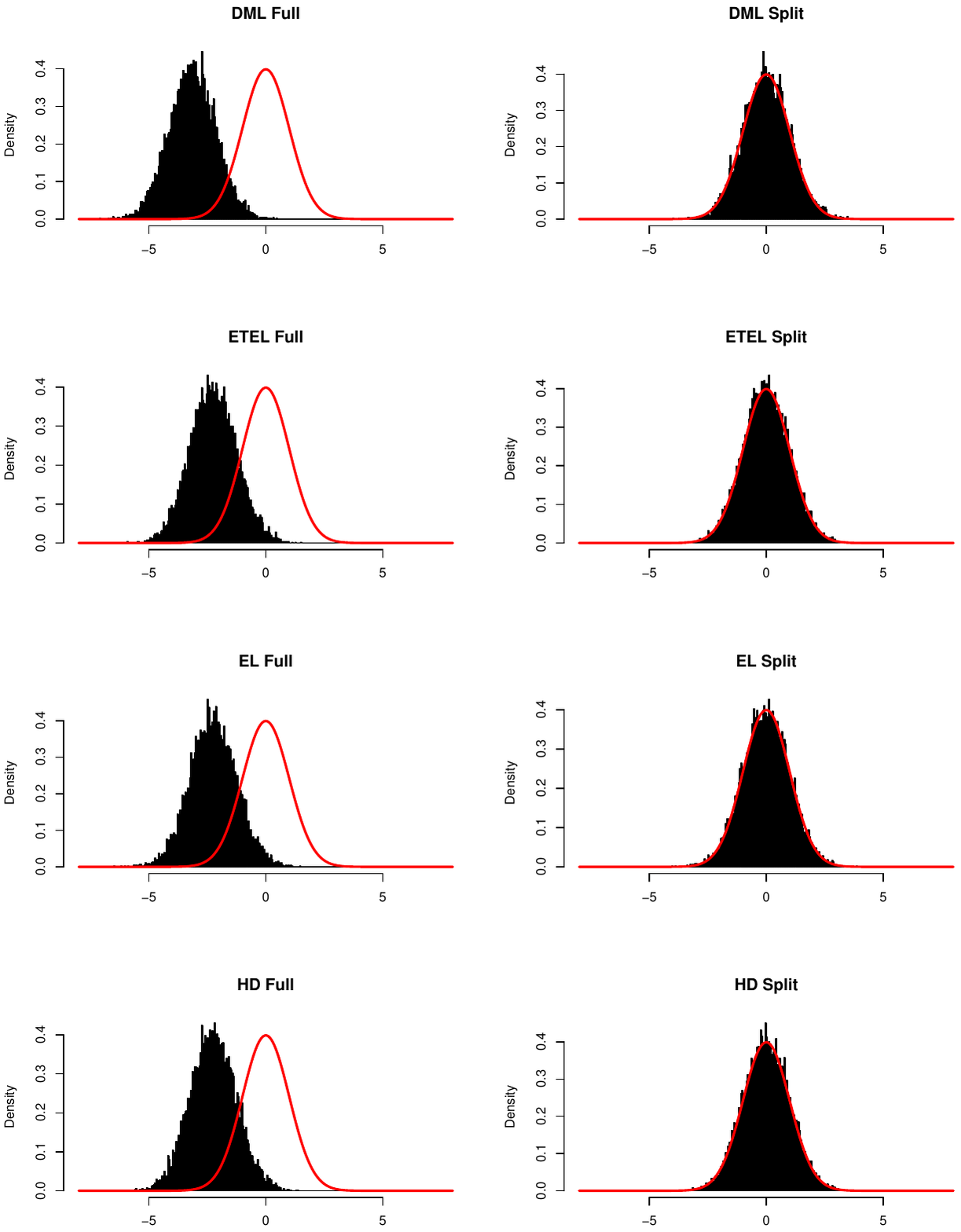}
	\caption{\label{fig:split} Comparison between double ML (DML) and Bayesian empirical likelihood methods using full-sample and sample splitting approaches ($n=500$). The red curve represents the standard normal density.}
\end{figure}

When using ML methods to estimate the nuisance functions, \cite{chernozhukov2018double} demonstrate that sample splitting plays a key role in reducing the bias when estimating $\beta$. When showing the consistency, the remainder term in the Taylor expansion involves the product between the error term in the PS model and the bias using ML method to estimate $\mu(\cdot)$. In some cases, this remainder term might not vanish when $n \to \infty$ as the two terms are correlated. In conventional semi-parametric analysis, we can impose Donsker conditions to restrict the class of functions that contains the estimator of $\mu(\cdot)$ so that the remainder term will be negligible. But when using ML methods where $p$ is modelled as $n$ increases, Donsker conditions are inappropriate. In Figure \ref{fig:split}, we replicate the example in \cite{chernozhukov2018double}, with $\hat \mu(X) = \mu(X) + (Y - \mu(X))/{n^{1/3}}$ and $\hat \pi(X) = \pi(X)$. We use the full sample to estimate $\beta$ using \eqref{plr} and Algorithm \ref{A1} with a vague prior $\mathcal{N}(0,10000)$ and $\lambda = 0,-1,-1/2$,  which corresponds to EL, ETEL and HD cases respectively. The first column of Figure \ref{fig:split} shows the results of 10,000 replicates with the histograms of the posterior mean via the Bayesian method in Algorithm \ref{A1}. In this simple example,  both methods indicate some biases from the full sample approach while the Bayesian method has a slightly smaller bias. To overcome this issue, \cite{chernozhukov2018double} propose the use of sample splitting, that is, the data are partitioned into $K$ groups.  The functions $\hat \mu_k(\cdot)$ and $\hat \pi_k(\cdot)$ are estimated using all the data excluding the $k$th group. Then the double ML estimator for $\beta$  is the solution to $ 1/K \sum_{k=1}^{K}\E_k [\psi(Z;\beta)]=0$, where $\E_k(\cdot)$ is the empirical expectation over the $k$th fold of the data. This creates the independence between two terms, leading to unbiased estimation. Therefore, it is necessary to amend Algorithm \ref{A1} to incorporate the sample splitting strategy to remove the bias. In essence, we can mimic the procedure to partition the data into $K$ groups and estimate the $\mu_k(\cdot)$ and $\pi_k(\cdot)$ using all the data excluding the $k$th group and then use them to obtain the non-parametric probability $p_i$ in \eqref{gel} with $i \in$ $k$th group only.  Algorithm \ref{A2} summarizes the  update algorithm to generate the posterior sample using this sample splitting procedure. 

\begin{algorithm}[H]
	\caption{\label{A2} Algorithm to obtain the posterior sample of $\beta$ via the Bayesian generalized empirical likelihood with sampling splitting.}
	\begin{algorithmic}[1]
		\Require{$\mathcal{D} = (z_1,\ldots,z_n)$}
		\State Partition the data into $K$ groups (roughly equal size), $\mathcal{D} = \left\{\mathcal{D}_1,\ldots,\mathcal{D}_K\right\}$.
		\For {$j$ \textbf{to} $1:J$}
		\For {$k$ \textbf{to} $1:K$}
		
		\State Estimate $\mu_k(\cdot)$ and $\pi_k(\cdot)$ using some ML methods with data $\mathcal{D}\setminus\mathcal{D}_k$.
		\State Obtain the non-parametric probability, $\left\{p^{(j)}_i\right\}_{i\in\mathcal{D}_k}$, by solving the optimization problem in \eqref{gel} with $\psi(z_i;\beta^{(j-1)})$ for $i\in\mathcal{D}_k$. 
		\EndFor
		\State	Sample $\beta^{(j)} \sim \pi_j (\beta \left|\mathbf{z}\right.)$ using the MCMC approach, 
		where $\pi_j(\beta\left|\mathbf{z}\right.)\propto \pi_0(\beta) \times  \prod_{i=1}^{n}p_i^{(j)}$.
		\EndFor
		\State \Return $\left(\beta^{(1)},\ldots,\beta^{(J)}\right)$.
	\end{algorithmic}
\end{algorithm}

The second column of Figure \ref{fig:split} displays the results using a two-fold sample splitting procedure, i.e., $K =2$. Both the double ML methods and the amended Bayesian approach effectively eliminate bias from the full-sample estimation procedure. The posterior distribution obtained using Algorithm \ref{A2} exhibits similar distributional properties across 10,000 replications.

\subsection{Validating the posterior uncertainty}
\label{sec:valid}

In this section, we evaluate whether the posterior in \eqref{postgel} is a valid posterior inference in terms of uncertainty. As demonstrated in \cite{chernozhukov2018double}, if the estimated functions of $\mu(\cdot)$ and $\pi(\cdot)$ converge to the true functions in probability under some regularity conditions, the estimator for $\beta$ in \eqref{plr1} will converge in distribution to a normal distribution, centered at $\beta$ with variance, $\E[(D-\pi(X))^2]^{-1} \E[(Y-\beta D-\mu(X))^2]$. This is the semi-parametric efficiency bound for $\beta$. 

The `posterior' density in Algorithm \ref{A2}, $\pi(\beta|z_{1:n})$, is a non-standard posterior inference as the likelihood is replaced by the profile likelihood with plug-in estimates for  $\mu(\cdot)$ and $\pi(\cdot)$. Despite the theoretical support of the ETEL case in \cite{schennach2007point,chib2018bayesian}, it remains important to assess whether the proposed Bayesian approach, combined with sample-splitting, yields valid Bayesian inference when the nuisance functions are estimated using ML methods.  That is, if interval $S_\alpha(z)$ is a designated $1-\alpha$ probability interval, a `posterior' density in Algorithm \ref{A2}, $\pi(\beta|z_{1:n})$, will have the property $\mathbb{P}\left[ \beta \in S_\alpha(z)) | z_{1:n}\right] = 1-\alpha$, if $Z_1,\ldots,Z_n$ are drawn from the true data generating model.  In particular, the proposed method incorporates cross-fitting together with Neyman-orthogonal moment conditions, which mitigate first-order bias arising from ML estimation of nuisance functions and relax the need for restrictive empirical process conditions, such as Donsker assumptions. However, it complicates the settings in \cite{schennach2007point, chib2018bayesian}. In particular, establishing posterior concentration and a Bernstein–von Mises theorem in this setting would require controlling how nuisance estimation error propagates through the inner optimization over implied probabilities in the generalized empirical likelihood construction. To our knowledge, such theoretical guarantees have not yet been developed. In the absence of a formal proof, we instead assess the calibration of the proposed posterior using the simulation-based diagnostic of \cite{monahan1992proper}.

\cite{monahan1992proper} proposed a notion of proper Bayesian inference by replacing the parametric likelihood with an alternative likelihood function. They stated that  the `posterior' density, which derives from the alternative likelihood, should follow the law of the probability deriving from the Bayes's rule. The posterior density is defined as valid by coverage if $\mathbb{P}_{\pi}\left[ \beta \in S_\alpha(z)) \right] = 1-\alpha$, if $Z_1,\ldots,Z_n$ are drawn from the true data generating model.  The posterior coverage set, $S_\alpha(z)$, resulting from {a valid posterior, should achieve nominal coverage under the joint measure of $Z$ and $\theta$}, that is, $P_{\pi} (\beta \in S_\alpha(z))$ should have expectation $1-\alpha$ for data generated under the measure  {$\pi_0(\beta) f(z|\beta)$ on $(\beta,Z)$} for every absolutely continuous prior, $\pi_0(\ldotp)$. To verify this property, let 
\begin{equation}
	\label{Hformula}
	H= \int_{-\infty} ^{\beta} \pi(\varphi|z) d \varphi, 
\end{equation}
and if $\pi(\beta|\mathbf{z})$ is a valid posterior, then $H$ follows Uniform$(0,1)$. In practice, if we generate $\beta_k (k=1,\ldots,m) \sim \pi_0(\ldotp)$ and the data, $z_{1:n}^{(k)}$, from  $f( \ldotp |\beta_k)$, and compute the posterior according to Algorithm \ref{A2}. Then we can obtain $H_k$ based on \eqref{Hformula} by replacing $\beta$ with $\beta_k$.   If the distribution of $H_k$ follows the uniform distribution, then the posterior distribution generated from Algorithm \ref{A2} is defined as a coverage proper posterior, yielding valid posterior inference.  This evaluation method for the validity of  posterior inference has also been applied to verify the correctness of Bayesian computation \citep[see for example,][]{talts2018validating}.

\begin{figure}[ht]
	\centering
	\includegraphics[scale=0.45]{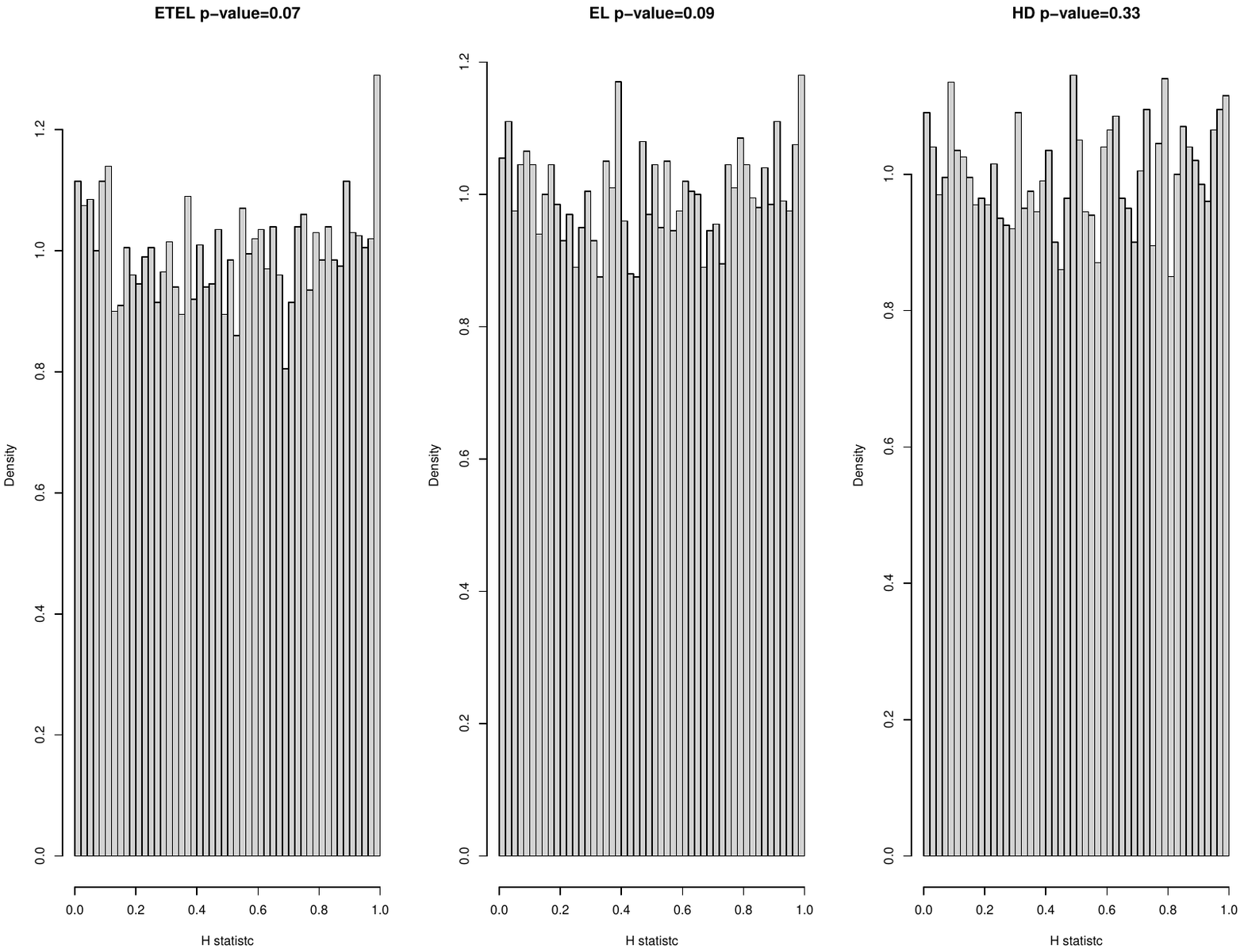}
	\caption{\label{gelplot} Histograms for $H$ statistics using the Bayesian generalized empirical likelihood. $P$-values represents the Kolmogorov–Smirnov test for uniformity of $H$.}
\end{figure}

In light of this approach, we investigate the validity of the proposed generalized empirical likelihood approach via a simulation study using the same set-up with the sample-splitting simulation, with each $\beta_k$ ($k=1,\ldots,10000)$  generating from $N(1,2)$. Figure \ref{gelplot} contains the histograms of the simulated $H$ over 10000 simulation runs with the associated $p$-values of the Kolmogorov–Smirnov test  for uniformity. For all cases, it suggests that simulation $H$ values follow uniform distribution and indicates proper posterior inference according to \cite{monahan1992proper} when $\lambda=0,-1,-1/2$. Therefore, this simulation result gives us  evidence that the proposed method according to Algorithm \ref{A2} can be regarded as a valid approach for posterior inference.

\section{Simulation}
\label{sec:sim}
In this section, we examine the performance of the proposed Bayesian method described in the previous section. We consider the following models for our simulation studies. 
\begin{itemize}
	\item EL, ETEL, HD: The Bayesian generalized empirical likelihood  with $\lambda=0,-1, -1/2$ respectively described in Section \ref{sec:BEL}, with calculation using Alogrithm \ref{A2}. 
	\item BDR-HD:  The Bayesian doubly robust high-dimension method proposed in \cite{antonelli2022causal}, where the propensity score and outcome are estimated via regression models with the GP prior, and then the MCMC estimate is plugged in to a doubly robust estimator. The variance is adjusted through the frequentist bootstrap so that it will achieve the nominal coverage rate.
	\item BDML: The Bayesian double machine learning approach proposed in \cite{ditraglia2025bayesian}, which based on the reduced-form system
		\[
		Y_i = X_i^\top \delta + \epsilon_i, \qquad
		D_i = X_i^\top \gamma + \varepsilon_i,
		\]
		where the error terms $(\epsilon_i,\varepsilon_i)$ are assumed jointly Gaussian with covariance matrix $\Sigma$. The causal parameter of interest is identified through the residual covariance structure after adjusting for the observed covariates $X_i$. For the prior specification, we adopt the BDML-Basic setup used in the simulation study of \cite{ditraglia2025bayesian}. We only consider this method in Example 2 due to the normality assumption of the treatment variable.
	\item DML: The frequentist double machine learning approach proposed in \cite{chernozhukov2018double}.
\end{itemize}
For all the Bayesian methods, we generate $5,000$ MCMC samples and $1,000$ burn-in iterations each simulation with $1,000$ simulation replicates.  The code for the simulation is publicly available on our GitHub page at \url{https://github.com/yumcgill/Bayesian-DML}.

\subsection{Binary exposure}

In this case, we consider $D$ is binary and simulate 
$$X=(X_1,X_2, \ldots,X_{p}) \sim \mathcal{N}_p\left(0,\Sigma\right), \Sigma_{ij} =
\begin{cases}
	1, & \text{if } i = j, \\
	0.3, & \text{if } i \neq j ,
\end{cases}$$ and then simulate
\begin{equation*}
	\begin{aligned}
		D\left|X \right. &\sim \text{ Bernoulli}\left(\text{expit}\left(0.3X_1+0.2X_2-0.4X_5\right)\right)\\
		Y \left|D,X\right. &\sim  \mathcal{N}\left(D+0.5X_1+X_3-0.1X_4-0.2X_7,1\right).\\
	\end{aligned}
\end{equation*}
We are interested in estimating the average treatment effect (ATE), i.e, $\beta$. In the analyses, we take $p=500$ and $n=50, 200$ and place a non-informative prior, $\mathcal{N}(0,10000)$, for the ATE. Table \ref{sim2s} shows the performance of ETEL, EL, HD, DML across ML approaches (lasso, random forest, and neural network) and BDR-HD in terms of bias, root mean squared error (RMSE), and coverage rate. For the proposed Bayesian method, it maintains low bias and RMSE with relatively high coverage rates. Specifically, EL and HD generally exhibit slightly better precision than ETEL, with HD often attaining the lowest bias.  In terms of ML approaches, neural networks, although effective in reducing bias, show higher RMSE and lower coverage rates compared to lasso and random forest. In addition, HD paired  with neural network achieves the lowest bias.  This pattern is consistent with the robustness properties of the HD criterion \citep{kitamura2001asymptotic}. In particular, when cross-fitted neural network nuisances introduce non-negligible second-order remainder terms, HD down-weights the influence of observations with extreme moment values more than ETEL, whose exponential tilting method inflates bias.  The DML method demonstrates quite similar results with the proposed Bayesian method, in line with our previous demonstration about the uncertainty quantification. BDR-HD shows similar bias as other methods, but it achieves the lowest RMSE and the highest coverage rate due to the extra bootstrap step to adjust the posterior variance. However, the running time per iteration for $n=50$ of BDR-HD is approximately five times longer than that of the proposed Bayesian method.  Overall, the proposed method demonstrates a reliable balance between the statistical performances and computational intensity.

\begin{table}[ht]
	\caption{\label{sim2s} Binary exposure:  Simulation results of the marginal causal effect under high-dimensional settings, with true value equal to 1, on 1000 simulation runs on generated datasets of size $n$. }
	\centering
	{		
		\begin{tabular}{cccc|ccc}
			\hline
			&\multicolumn{3}{c}{$n=50$} & \multicolumn{3}{c}{$n=200$}\\
			& Bias & RMSE & Coverage rate (\%) & Bias & RMSE & Coverage rate (\%) \\
			\hline
			ETEL (Lasso)&0.08 &0.42   &93.1  & 0.08&0.19 &93.4   \\
			EL (Lasso)& 0.07& 0.42&91.7 & 0.07 & 0.19  &92.1  \\
			HD (Lasso)&0.08&  0.42 & 92.8 & 0.04 & 0.18& 91.8  \\
			ETEL (Random forest)& 0.06&  0.43 & 90.9 & 0.08 & 0.19 & 92.4 \\
			EL (Random forest)&0.06 & 0.42& 92.8 & 0.07 & 0.20 & 92.0 \\
			HD (Random forest)&0.07&  0.43 & 90.9 & 0.08 & 0.19 & 92.7 \\
			ETEL (Neural network)&  0.08& 0.47    & 87.6 & 0.03&  0.22&87.4  \\
			EL (Neural network)&0.06 & 0.44 & 90.5 & 0.04 & 0.23  & 88.3 \\
			HD (Neural network)& 0.01 & 0.46   & 90.7  & 0.02  &0.22 &90.0\\
			DML (Lasso) & 0.10 &0.41 &92.7 & 0.05 & 0.17 &93.1\\
			DML (Random forest) & 0.09 & 0.41 & 93.9 & 0.08& 0.20 &92.0\\
			DML (Neural network)& 0.08 & 0.43 & 90.2 &0.05 & 0.22 & 88.8\\
			BDR-HD &0.09 & 0.36 &97.7  &0.07&0.15&95.4\\
			\hline
		\end{tabular}
	}
\end{table}

\subsection{Continuous exposure}
In this example, we consider $D$ is continuous and simulate $X=(X_1,X_2, \ldots,X_{p}) \sim \mathcal{N}_p\left(0,\Sigma\right)$ and $\Sigma_{ij} =1$ if $i=j$ and 0.05 otherwise, and then simulate
\begin{equation*}
	\begin{aligned}
		D\left|X \right. &\sim \mathcal{N}\left(0.45X_1+0.9X_2-0.4X_5, 1\right)\\
		Y \left|D,X\right. &\sim  \mathcal{N}\left(D+0.5X_1+X_3-0.1X_4-0.2X_7,1\right).\\
	\end{aligned}
\end{equation*}
In this example, the coefficient associated with $D$ is the main regression coefficient and is the parameter of interest. We set $p=40$ and $n=40$, and in the proposed Bayesian method, we assign a relatively informative prior, $\mathcal{N}(1,2)$, to $\beta$. Table \ref{sim2c} shows the results of bias, RMSE and coverage rate across all methods. Most of the methods perform well in terms of bias, ranging between $-0.01$ and $0.06$, while the RMSE varies slightly, with most values in between  $0.13$ and $0.19$. In particular, Lasso-based approaches have very low bias ($0.01-0.02$), while random forest based methods show slightly higher bias. However, BDML displays noticeable bias due to small sample size relative to the number of covariates $p$, reflecting the difficulty of accurately estimating nuisance components in high-dimensional settings with limited observations. In terms of RMSE, the proposed Bayesian method displays smallest RMSEs $(0.13-0.14)$ across all ML methods. DML methods showing a slightly higher RMSE $(0.16-0.19)$, while BDR-HD has a much higher RMSE. 
This is mainly because  BDR-HD uses an estimator that involves inverse weighting by the estimated generalized propensity score (GPS).  For units whose observed exposure value is far from their predicted exposure, this conditional density can be extremely small, and dividing by a near-zero density value produces extreme weights that inflate variance dramatically. The proposed Bayesian methods and DML avoid this issue entirely because they are based on the Neyman orthogonal score equation, which does not require weighting by the propensity score. In practice, if inverse-weighting approaches are used, additional measures such as exposure binning to mitigate variability \citep{naimi2014constructing} or the use of a stabilized GPS should be considered. Regarding coverage rates, EL-based methods always achieve close to nominal level rates $(90.6-94.0)$, while ETEL-based methods show a bit lower coverage $(85.8-89.9)$. BDR-HD reaches the coverage rate at 95.9, closest to the nominal level. We also notice that methods using neural network generally have lower coverage rates than methods using Lasso and random forest. Results indicate that the proposed Bayesian method with EL-based likelihood, coupled with an informative prior,  performs well overall in terms of balancing low bias and adequate coverage rates. 

\begin{table}[ht]
	\caption{\label{sim2c} Continuous exposure:  Simulation results of the marginal causal effect under high-dimensional settings, with true value equal to 1, on 1000 simulation runs with $n=40$ and $p=40$.}
	\centering
	{		
		\begin{tabular}{cccc}
			\hline
			& Bias & RMSE & Coverage rate (\%) \\
			\hline
			ETEL (Lasso)&0.02&  0.13 &   89.6  \\
			EL (Lasso)& 0.02& 0.13&94.0 \\
			HD (Lasso)&0.01&  0.13 & 91.7    \\
			ETEL (Random forest)& 0.06& 0.13 & 89.9  \\
			EL (Random forest)& 0.06& 0.13&93.0\\
			HD (Random forest)& 0.06&0.13 & 91.5 \\
			ETEL (Neural network)&0.04  &0.14   &85.8    \\
			EL (Neural network)& 0.04& 0.14 & 90.6  \\
			HD (Neural network)& 0.04 & 0.14  &  88.0  \\
			DML (Lasso) & -0.01 &0.19 &90.2 \\
			DML (Random forest) & 0.02& 0.16&92.3 \\
			DML (Neural network)& 0.01 & 0.18& 88.5\\
			BDR-HD & 0.03 & 0.75 & 95.9  \\
			BDML& -0.92 & 1.05 & 88.0\\
			\hline
		\end{tabular}
	}
\end{table}

\section{Real data application}
\label{sec:app}
In this section, we apply the proposed methodology to the Stop and Search data, described in Section \ref{data} in London to enhance our understanding of the impact of the racial disproportionality, particularly on expressive crimes. To conduct an informative  analysis, we assign an informative prior to the model, specifically $\mathcal{N}(0,2)$, which reflects our initial assumption that there is no effect of disproportionality in stop and search practices. In terms of confounders, we include variables such as percentage of males (gender), unemployment number and migrant rate,  with a total of 31 variables. A summary of all variables, including their descriptive statistics aggregated in the borough level in the analysis, is given in Table  \ref{tab:Descriptive statistics of variables}. We estimate the both $\pi(\cdot)$ and $\mu(\cdot)$ using all the confounders listed in Table \ref{tab:Descriptive statistics of variables}, and then apply these estimates in the Neyman-orthogonal equation via the Bayesian generalized empirical likelihood. We implement the Bayesian generalized empirical likelihood  with $\lambda=0,-1, -1/2$ coupled with Lasso, random forest and neural network methods, and generate in total 10,000 posterior samples in each case with 1,000 burn-in iterations. For comparison, we also include DML, BDR-HD and BDML methods used in the simulation study. Table \ref{res} reports the estimation results across different methods. Across all cases, the posterior mean and the point estimates are consistently negative. For the proposed method, while there is little variation across the ETEL, EL, and HD metrics, the choice of ML method has a more noticeable impact on the width of the confidence intervals.  Those for Lasso are the widest, and this is due to the high correlation among some features, leading to high variability in cross-validation when selecting the optimal tuning parameter. Random forest and neural network models have relatively narrower confidence intervals, with the narrowest under EL and HD methods among them. The DML estimates are also negative across all specifications, with smaller magnitudes ranging from -0.35 to -0.16, and relatively tight confidence intervals. The BDML estimator produces a negative posterior mean as well, although with greater posterior uncertainty. Moreover, the BDR-HD estimator yields the most negative estimate. This comparatively large magnitude may reflect the estimator’s sensitivity to the doubly robust weighting scheme, as observed in the simulation study.

Overall, all methods suggest a negative relationship in terms of the posterior mean or point estimate, indicating that accounting for the proportion of the Black population can help alleviate disproportionality in stop and search practices targeted at the Black community. Specifically, when focusing on stop and search incidents involving Black individuals for expressive crimes, we observe that as the percentage of Black residents in a borough increases, the predicted value of the DI for these practices decreases. In other words, as the Black population share increases, the relative disparity between the observed stop and search rate for Black individuals and the London-wide benchmark becomes smaller. As a result, these boroughs exhibit lower levels of measured disproportionality, indicating that stop and search activity involving Black individuals is more proportionate to the local demographic composition relative to boroughs with smaller Black populations. This negative association is consistent with prior research indicating that policing intensity is shaped by socioeconomic condition and local demographics, with police officers tending to conduct more searches in more economically unequal areas \citep{suss2023}. In addition, the negative effect should be interpreted with care. A declining DI does not necessarily imply a reduction in the absolute level of stop and search activity targeting Black individuals, but rather a reduction in relative disproportionality compared to the London-wide benchmark. This distinction is important, as boroughs with lower DI values may still experience high levels of policing intensity in absolute terms.

\begin{table}[ht]
	\centering
	\caption{Descriptive statistics of the Stop and Search data in London}
	\label{tab:Descriptive statistics of variables}
	\footnotesize
	\begin{tabular}{p{0.5\textwidth}llll}
		\hline
		Variable & Min. & Max. & Mean & Std. Dev. \\
		\hline
		Dispropotionality index & 1.66 &12.58 & 4.12 & 2.46 \\
		Proportion of Black population & 1.89 & 17.59 & 12.64 & 6.92 \\
		Population Density & 2198 & 15703 & 7291 & 3670.89 \\
		Proportion of Male Residents & 0.47 & 0.55 & 0.49 & 0.01 \\
		Proportion of Migrant Residents & 0.57 & 6.42 & 1.94 & 1.30 \\
		Number of unemployment & 2653 & 123179 & 83149 & 24860.92 \\
		Proportion of Residents Self-claimed as Unhealthy &2.72 & 5.55 & 4.23 &0.63 \\
		Proportion of Disabled Residents & 21.41 & 32.38 & 26.41 & 2.37\\
		Proportion of Students Residents & 13.90 & 28.54 & 21.90 &2.53 \\
		Proportion of Households without Cars & 21.53 & 77.20 & 42.90 & 16.71 \\
		Proportion of Households in Renting & 29.54 & 74.26 & 53.37 & 13.84 \\
		Household Room Occupation Rate & 9.39 & 45.02 &27.40 & 7.62 \\
		Proportion of Residents with Higher Education Degree & 29.52 & 74.18 & 47.95 & 9.93 \\
		Proportion of Deprived Households & 38.96 & 62.41 & 51.46 & 5.45\\
		Proportion of Residents Lacking Care Supports & 91.30 & 93.70 & 92.29  & 0.62\\
		Proportion of Households Lack of Family Cohension & 33.01 & 54.54 & 42.87 &  5.61\\
		Proportion of Households Living in Unstable Status & 13.93 & 43.27 & 30.55 & 6.34\\
		Proportion of Greenspace Area & 0.01 & 0.487 & 0.17 & 0.12\\
		Proportion of Young Residents Under 18 & 0.28 & 0.49 & 0.37 &  0.05\\
		Areas (km$^2$) & 2.90 & 150.14 & 47.68 & 32.75 \\
		Density of Roads & 61.23 & 258.62 & 128.08 &  41.78\\         
		Density of Manufacturing Stores and Places & 3.00 & 94.00 & 13.82 &16.15 \\
		Number of Residential Places & 420 & 3816 & 1094 &  603.35\\
		Density of Residential Places & 6.00 & 401.00 & 47.88 & 73.89 \\       
		Number of Manufacturing Stores and Places & 195 & 730 & 401 & 127.83\\  
		Number of Public Transport Stations & 282.00 & 1821.00& 957.40 &  353.16\\
		Density of Public Transport Stations & 10.00 & 97.00 & 13.82 &  16.02\\
		Number of Pubs & 28.00 & 447.00& 123.10 & 82.46 \\  
		Density of Pubs & 0.62 & 76.10& 6.37 & 13.39 \\  
		Number of Retail Stores & 486 & 3536 & 1282 & 522.86\\
		Density of Retail Stores & 8.00 & 167.00 & 44.94& 42.29\\
		Number of Schools & 8.00 & 171.00 & 102.20 & 29.56 \\
		Density of Schools & 0.85 & 8.00 & 2.98& 1.83 \\
		\hline
	\end{tabular}
\end{table}

\begin{table}[ht]
	\caption{\label{res} Posterior mean/point estimate (estimate) of the effect of expressive crime involving the Black population, along with the associated 95\% credible/confidence interval (C.I.).}
	\centering
	{		
		\begin{tabular}{ccc}
			\hline
			& Estimate &  95\% C.I.\\
			\hline
			ETEL (Lasso)&$-0.59$ &   $(-2.17, 0.95)$  \\
			EL (Lasso)& $-0.56$& $(-2.21, 0.99)$\\
			HD (Lasso)&$-0.55$&  $(-2.18, 1.00)$   \\
			ETEL (Random forest)&  $-0.31$& $(-0.52, -0.14)$  \\
			EL (Random forest)& $-0.31$& $(-0.53, -0.14)$\\
			HD (Random forest)& $-0.31$&$(-0.55, -0.14)$ \\
			ETEL (Neural network)&$-0.41$ &$(-0.73, -0.16)$   \\
			EL (Neural network)& $-0.40$& $(-0.78, -0.14)$  \\
			HD (Neural network)& $-0.41$	 & $(-0.78, -0.16)$ \\
			DML (Lasso) & $-0.16$ & $(-0.26, -0.07)$  \\
			DML (Neural network)& $-0.21$ & $(-0.36, -0.06)$ \\
			DML (Trees)& $-0.22$ &$(-0.32, -0.11)$ \\
			DML (Boosting)& $-0.35$ & $(-0.57, -0.14)$ \\
			BDR-HD & $-1.65$ & $(-2.13, -1.16)$  \\
			BDML & $-0.14$ & $(-1.39, 0.44)$\\
			\hline
		\end{tabular}
	}
\end{table}

\section{Discussion}
\label{sec:dis}
In this paper, we introduce a Bayesian procedure for semi-parametric partially linear structural regression which allows us to place a specific prior to the parameter of interest. In addition, this approach also addresses issues in high-dimensional scenarios. By integrating double ML method and sample splitting procedure into our Bayesian framework, it preserves key asymptotic properties and consistent estimation on the parameter of interest.  In particular, we showed that computations following this paradigm yield valid posterior inference in terms of coverage  according to \cite{monahan1992proper}. Our approach also accommodates flexible ML models for treatment and outcome, mitigating the impact of model misspecification. Although we focus on a partially linear model in this paper, the proposed framework can be extended to a broad class of orthogonal moment conditions demonstrated in \cite{chernozhukov2018double}.  In particular, any causal estimand that admits a Neyman-orthogonal score representation can, in principle, be incorporated within our Bayesian double ML framework. This includes a range of targets such as average treatment effects and heterogeneous treatment effects, under binary, continuous, or multivalued treatments. Extending the framework to these settings would primarily involve specifying the corresponding orthogonal score and adapting the proposed posterior construction. It preserves the advantages of flexible nuisance estimation using ML methods and coherent uncertainty quantification.

In our application, we estimate the effect of impact of ethnic disproportionality in stop and search for expressive crimes in London with an informative prior. Our results suggest that stop and search practices are disproportionately distributed across London boroughs, and that disproportionality involving the Black community is lower in boroughs with higher proportions of Black residents. In other words, our analysis suggests that stop and search practices involving Black individuals are relatively more disproportionate in areas that are predominantly white. The estimand considered in our application differs from previous estimands, such as the individual-level estimand proposed by \cite{zhao2022note}, which contrasts outcomes for minority and non-minority individuals. The outcome of interest in our analysis is a standardized areal-level count of stop and search events, rather than the use of force following an individual-level stop interaction or detainment, as in \cite{knox2020administrative,zhao2022note}. Our DI compares the intensity of stop interactions involving Black individuals across different boroughs of London. In this sense, we are not examining whether Black individuals are treated worse than individuals of other races, but rather how the intensity of stop and search involving Black individuals varies geographically. This distinction is important for interpreting the DI as evidence of potential discrimination. Elevated or spatially patterned DI values may indicate potential structural or geographic disproportionality in policing, and may therefore help identify boroughs where stop and search practices warrant closer scrutiny. However, because the estimand is defined at the areal level and does not directly contrast individual-level outcomes across racial groups, it should not be interpreted as definitive evidence of individual-level racial discrimination. It could be, for example, that racial bias is severe in terms of how minority groups are treated compared to non-minority groups, but that this bias is distributed relatively evenly across the city; such a pattern would not necessarily be captured by our estimand.

A future analysis involving individual-level data comprising different ethnic groups could entail estimating the causal risk ratios discussed in \cite{zhao2022note} using the proposed Bayesian double ML methodology, leveraging a linearization of the estimand as in \cite{jiang2025double}. The main advantage of the Bayesian double ML framework in this case would be its ability to combine flexible ML for nuisance estimation with posterior summaries for the causal risk ratio, especially in settings where external data sources are integrated and multiple nuisance functions must be estimated. This may be useful in discrimination studies where researchers wish to incorporate prior information, assess sensitivity to different priors, or make direct probabilistic statements about racial disparities.

It is also worth mentioning that the empirical analysis of the DI does not take into account the spatial dimension of the data. While this issue is outside the scope of this work, it is possible that there are unobserved spatial confounders. In future work, the proposed framework could be extended to address the threat of spatial confounding either through distance-adjusted propensity matching as in \cite{papadogeorgou2019adjusting} or through a probabilistic representation of the latent confounders similar to \cite{papadogeorgou2023spatial}. Moreover, we should note that residual spatial autocorrelation was not accounted for in this application. Another important future extension would involve incorporating a Gaussian random Markov field component to the model to account for spatial dependence.

Bayesian methods hold significant interest in applied scientific causal research. These methods enable direct probability statements regarding treatment effectiveness and facilitate sensitivity assessments with different prior or expert inputs. There remains many scopes for future research to explore various possibilities in Bayesian semi-parametric inference in high-dimensional settings. An important future direction is the development of formal theoretical guarantees for the proposed methodology as we discussed in Section \ref{sec:valid}. In addition, combining Bayesian methods with modern ML tools offers considerable potential for handling increasingly complex data structures in causal inference, including time-varying treatments, survival data, interference settings, network dependence, and hierarchical or multilevel models in high-dimensional applications.

\section*{Acknowledgments}
This research has been supported by the Engineering \& Physical Sciences Research Council (EP/Y029755/1). The authors would like to thank the Evidence and Insight team from MOPAC (Mayor's Office for Policing and Crime, Greater London Authority) for their valuable comments and suggestions that contributed to the progress of this research.

\vspace{0.2in}
\bibliographystyle{chicago}
\bibliography{../StopSearch}

\begin{thebibliography}{}

\bibitem[\protect\citeauthoryear{Antonelli, Papadogeorgou, and
  Dominici}{Antonelli et~al.}{2022}]{antonelli2022causal}
Antonelli, J., G.~Papadogeorgou, and F.~Dominici (2022).
\newblock Causal inference in high dimensions: A marriage between {B}ayesian
  modeling and good frequentist properties.
\newblock {\em Biometrics\/}~{\em 78\/}(1), 100--114.

\bibitem[\protect\citeauthoryear{Baggerly}{Baggerly}{1998}]{baggerly1998empirical}
Baggerly, K.~A. (1998).
\newblock Empirical likelihood as a goodness-of-fit measure.
\newblock {\em Biometrika\/}~{\em 85\/}(3), 535--547.

\bibitem[\protect\citeauthoryear{Bong and Lee}{Bong and
  Lee}{2023}]{bong2023local}
Bong, S. and K.~Lee (2023).
\newblock Local causal effects with continuous exposures: A matching estimator
  for the average causal derivative effect.
\newblock {\em arXiv preprint arXiv:2311.18532\/}.

\bibitem[\protect\citeauthoryear{Bowling and Phillips}{Bowling and
  Phillips}{2007}]{bowling2007disproportionate}
Bowling, B. and C.~Phillips (2007).
\newblock Disproportionate and discriminatory: Reviewing the evidence on police
  stop and search.
\newblock {\em The Modern Law Review\/}~{\em 70\/}(6), 936--961.

\bibitem[\protect\citeauthoryear{Chamberlain and Imbens}{Chamberlain and
  Imbens}{2003}]{chamberlain2003nonparametric}
Chamberlain, G. and G.~W. Imbens (2003).
\newblock Nonparametric applications of {B}ayesian inference.
\newblock {\em Journal of Business \& Economic Statistics\/}~{\em 21\/}(1),
  12--18.

\bibitem[\protect\citeauthoryear{Chernozhukov, Chetverikov, Demirer, Duflo,
  Hansen, Newey, and Robins}{Chernozhukov
  et~al.}{2018}]{chernozhukov2018double}
Chernozhukov, V., D.~Chetverikov, M.~Demirer, E.~Duflo, C.~Hansen, W.~Newey,
  and J.~Robins (2018).
\newblock Double/debiased machine learning for treatment and structural
  parameters.
\newblock {\em The Econometrics Journal\/}~{\em 21\/}(1), C1--C68.

\bibitem[\protect\citeauthoryear{Chernozhukov and Hong}{Chernozhukov and
  Hong}{2003}]{chernozhukov2003mcmc}
Chernozhukov, V. and H.~Hong (2003).
\newblock An {MCMC} approach to classical estimation.
\newblock {\em Journal of Econometrics\/}~{\em 115\/}(2), 293--346.

\bibitem[\protect\citeauthoryear{Chib, Shin, and Simoni}{Chib
  et~al.}{2018}]{chib2018bayesian}
Chib, S., M.~Shin, and A.~Simoni (2018).
\newblock Bayesian estimation and comparison of moment condition models.
\newblock {\em Journal of the American Statistical Association\/}~{\em
  113\/}(524), 1656--1668.

\bibitem[\protect\citeauthoryear{DiTraglia and Liu}{DiTraglia and
  Liu}{2025}]{ditraglia2025bayesian}
DiTraglia, F.~J. and L.~Liu (2025).
\newblock Bayesian double machine learning for causal inference.
\newblock {\em arXiv preprint arXiv:2508.12688\/}.

\bibitem[\protect\citeauthoryear{Gaebler, Cai, Basse, Shroff, Goel, and
  Hill}{Gaebler et~al.}{2022}]{gaebler2022causal}
Gaebler, J., W.~Cai, G.~Basse, R.~Shroff, S.~Goel, and J.~Hill (2022).
\newblock A causal framework for observational studies of discrimination.
\newblock {\em Statistics and Public Policy\/}~{\em 9\/}(1), 26--48.

\bibitem[\protect\citeauthoryear{Graham, McCoy, and Stephens}{Graham
  et~al.}{2016}]{graham2016approximate}
Graham, D.~J., E.~J. McCoy, and D.~A. Stephens (2016).
\newblock Approximate {B}ayesian inference for doubly robust estimation.
\newblock {\em Bayesian Analysis\/}~{\em 11\/}(1), 47--69.

\bibitem[\protect\citeauthoryear{Hahn}{Hahn}{1998}]{hahn1998role}
Hahn, J. (1998).
\newblock On the role of the propensity score in efficient semiparametric
  estimation of average treatment effects.
\newblock {\em Econometrica\/}, 315--331.

\bibitem[\protect\citeauthoryear{Hirano and Imbens}{Hirano and
  Imbens}{2004}]{hirano2004propensity}
Hirano, K. and G.~Imbens (2004).
\newblock The propensity score with continuous treatments. applied bayesian
  modeling and causal inference from incomplete-data perspectives.

\bibitem[\protect\citeauthoryear{{HM Inspectorate of Constabulary and Fire \&
  Rescue Services}}{{HM Inspectorate of Constabulary and Fire \& Rescue
  Services}}{2022}]{hminspec2022}
{HM Inspectorate of Constabulary and Fire \& Rescue Services} (2022).
\newblock State of policing: The annual assessment of policing in england and
  wales 2022.
\newblock {\em HMICFRS\/}.

\bibitem[\protect\citeauthoryear{Huang, Beck, and Antonelli}{Huang
  et~al.}{2024}]{huang2024causal}
Huang, Z., B.~Beck, and J.~Antonelli (2024).
\newblock Causal inference and racial bias in policing: New estimands and the
  importance of mobility data.
\newblock {\em arXiv:2409.08059\/}.

\bibitem[\protect\citeauthoryear{Imbens, Johnson, and Spady}{Imbens
  et~al.}{1998}]{imbens1998information}
Imbens, G., P.~Johnson, and R.~H. Spady (1998).
\newblock Information theoretic approaches to inference in moment condition
  models.
\newblock {\em Econometrica\/}~{\em 66\/}(2), 333--357.

\bibitem[\protect\citeauthoryear{Jiang, Talbot, Carazo, and Schnitzer}{Jiang
  et~al.}{2025}]{jiang2025double}
Jiang, C., D.~Talbot, S.~Carazo, and M.~E. Schnitzer (2025).
\newblock A double machine learning approach for the evaluation of {COVID}-19
  vaccine effectiveness under the test-negative design: Analysis of
  {Q}u{\'e}bec administrative data.
\newblock {\em Statistics in Medicine\/}~{\em 44\/}(5), e70025.

\bibitem[\protect\citeauthoryear{Jung, Corbett-Davies, Gaebler, Shroff, and
  Goel}{Jung et~al.}{2018}]{jung2018mitigating}
Jung, J., S.~Corbett-Davies, J.~D. Gaebler, R.~Shroff, and S.~Goel (2018).
\newblock Mitigating included-and omitted-variable bias in estimates of
  disparate impact.
\newblock {\em arXiv:1809.05651\/}.

\bibitem[\protect\citeauthoryear{Kaplan and Chen}{Kaplan and
  Chen}{2012}]{kaplan2012two}
Kaplan, D. and J.~Chen (2012).
\newblock A two-step {B}ayesian approach for propensity score analysis:
  Simulations and case study.
\newblock {\em Psychometrika\/}~{\em 77\/}(3), 581--609.

\bibitem[\protect\citeauthoryear{Kitamura}{Kitamura}{2001}]{kitamura2001asymptotic}
Kitamura, Y. (2001).
\newblock Asymptotic optimality of empirical likelihood for testing moment
  restrictions.
\newblock {\em Econometrica\/}~{\em 69\/}(6), 1661--1672.

\bibitem[\protect\citeauthoryear{Kitamura, Otsu, and Evdokimov}{Kitamura
  et~al.}{2013}]{kitamura2013robustness}
Kitamura, Y., T.~Otsu, and K.~Evdokimov (2013).
\newblock Robustness, infinitesimal neighborhoods, and moment restrictions.
\newblock {\em Econometrica\/}~{\em 81\/}(3), 1185--1201.

\bibitem[\protect\citeauthoryear{Knox, Lowe, and Mummolo}{Knox
  et~al.}{2020}]{knox2020administrative}
Knox, D., W.~Lowe, and J.~Mummolo (2020).
\newblock Administrative records mask racially biased policing.
\newblock {\em American Political Science Review\/}~{\em 114\/}(3), 619--637.

\bibitem[\protect\citeauthoryear{Lee}{Lee}{2018}]{lee2018simple}
Lee, M.-J. (2018).
\newblock Simple least squares estimator for treatment effects using propensity
  score residuals.
\newblock {\em Biometrika\/}~{\em 105\/}(1), 149--164.

\bibitem[\protect\citeauthoryear{Liu, Saarela, Feldman, and Pullenayegum}{Liu
  et~al.}{2020}]{liu2020estimation}
Liu, K., O.~Saarela, B.~M. Feldman, and E.~Pullenayegum (2020).
\newblock Estimation of causal effects with repeatedly measured outcomes in a
  {B}ayesian framework.
\newblock {\em Statistical Methods in Medical Research\/}~{\em 29\/}(9),
  2507--2519.

\bibitem[\protect\citeauthoryear{Luo, Graham, and McCoy}{Luo
  et~al.}{2023}]{luoemp2021}
Luo, Y., D.~J. Graham, and E.~J. McCoy (2023).
\newblock Semiparametric {B}ayesian doubly robust causal estimation.
\newblock {\em Journal of Statistical Planning and Inference\/}~{\em 225},
  171--187.

\bibitem[\protect\citeauthoryear{Luo, Stephens, Graham, and McCoy}{Luo
  et~al.}{2023}]{luo2021bayesian}
Luo, Y., D.~A. Stephens, D.~J. Graham, and E.~J. McCoy (2023).
\newblock Assessing the validity of {B}ayesian inference using loss functions.
\newblock {\em arXiv:2103.04086\/}.

\bibitem[\protect\citeauthoryear{McCandless, Douglas, Evans, and
  Smeeth}{McCandless et~al.}{2010}]{mccandless2010cutting}
McCandless, L.~C., I.~J. Douglas, S.~J. Evans, and L.~Smeeth (2010).
\newblock Cutting feedback in {B}ayesian regression adjustment for the
  propensity score.
\newblock {\em The International Journal of Biostatistics\/}~{\em 6\/}(2), 16.

\bibitem[\protect\citeauthoryear{Meng}{Meng}{2017}]{meng2017}
Meng, Y. (2017).
\newblock {Profiling minorities: Police stop-and-search practices in
  contemporary London}.
\newblock {\em Human Geographies\/}~{\em 11\/}(1), 5--22.

\bibitem[\protect\citeauthoryear{Millner}{Millner}{2020}]{miller2020}
Millner, N. (2020).
\newblock As the drone flies: Configuring a vertical politics of urban
  policing.
\newblock {\em Political Geography\/}~{\em 80}, 102163.

\bibitem[\protect\citeauthoryear{Monahan and Boos}{Monahan and
  Boos}{1992}]{monahan1992proper}
Monahan, J.~F. and D.~D. Boos (1992).
\newblock Proper likelihoods for {B}ayesian analysis.
\newblock {\em Biometrika\/}~{\em 79\/}(2), 271--278.

\bibitem[\protect\citeauthoryear{{MOPAC}}{{MOPAC}}{2023}]{mopac2023}
{MOPAC} (2023).
\newblock Disproportionality board data pack.
\newblock {\em Disproportionality Board Data Pack 2023\/}.

\bibitem[\protect\citeauthoryear{Naimi, Moodie, Auger, and Kaufman}{Naimi
  et~al.}{2014}]{naimi2014constructing}
Naimi, A.~I., E.~E. Moodie, N.~Auger, and J.~S. Kaufman (2014).
\newblock Constructing inverse probability weights for continuous exposures: a
  comparison of methods.
\newblock {\em Epidemiology\/}~{\em 25\/}(2), 292--299.

\bibitem[\protect\citeauthoryear{Newey and Smith}{Newey and
  Smith}{2004}]{newey2004higher}
Newey, W.~K. and R.~J. Smith (2004).
\newblock {Higher order properties of GMM and generalized empirical likelihood
  estimators}.
\newblock {\em Econometrica\/}~{\em 72\/}(1), 219--255.

\bibitem[\protect\citeauthoryear{Oberwittler and Roché}{Oberwittler and
  Roché}{2022}]{ober2022}
Oberwittler, D. and S.~Roché (2022).
\newblock {How institutional contexts shape police-adolescent relations in
  France and Germany: Spatial and social disparities}.
\newblock {\em Policing and Society\/}~{\em 32\/}(3), 378--410.

\bibitem[\protect\citeauthoryear{Owen}{Owen}{2001}]{owen2001empirical}
Owen, A.~B. (2001).
\newblock {\em Empirical likelihood}.
\newblock CRC press.

\bibitem[\protect\citeauthoryear{Papadogeorgou, Choirat, and
  Zigler}{Papadogeorgou et~al.}{2019}]{papadogeorgou2019adjusting}
Papadogeorgou, G., C.~Choirat, and C.~M. Zigler (2019).
\newblock Adjusting for unmeasured spatial confounding with distance adjusted
  propensity score matching.
\newblock {\em Biostatistics\/}~{\em 20\/}(2), 256--272.

\bibitem[\protect\citeauthoryear{Papadogeorgou and Samanta}{Papadogeorgou and
  Samanta}{2023}]{papadogeorgou2023spatial}
Papadogeorgou, G. and S.~Samanta (2023).
\newblock Spatial causal inference in the presence of unmeasured confounding
  and interference.
\newblock {\em arXiv preprint arXiv:2303.08218\/}.

\bibitem[\protect\citeauthoryear{Qin and Lawless}{Qin and
  Lawless}{1994}]{qin1994empirical}
Qin, J. and J.~Lawless (1994).
\newblock Empirical likelihood and general estimating equations.
\newblock {\em The Annals of Statistics\/}~{\em 22\/}(1), 300--325.

\bibitem[\protect\citeauthoryear{Read and Cressie}{Read and
  Cressie}{2012}]{read2012goodness}
Read, T.~R. and N.~A. Cressie (2012).
\newblock {\em Goodness-of-fit statistics for discrete multivariate data}.
\newblock Springer Science \& Business Media.

\bibitem[\protect\citeauthoryear{Robinson}{Robinson}{1988}]{robinson1988root}
Robinson, P.~M. (1988).
\newblock Root-$n$-consistent semiparametric regression.
\newblock {\em Econometrica\/}~{\em 56\/}(4), 931--954.

\bibitem[\protect\citeauthoryear{Rosenbaum and Rubin}{Rosenbaum and
  Rubin}{1983}]{rosenbaum1983central}
Rosenbaum, P.~R. and D.~B. Rubin (1983).
\newblock The central role of the propensity score in observational studies for
  causal effects.
\newblock {\em Biometrika\/}~{\em 70\/}(1), 41--55.

\bibitem[\protect\citeauthoryear{Rubin}{Rubin}{1981}]{rubin1981bayesian}
Rubin, D.~B. (1981).
\newblock {The Bayesian bootstrap}.
\newblock {\em The Annals of Statistics\/}~{\em 9\/}(1), 130--134.

\bibitem[\protect\citeauthoryear{Schennach}{Schennach}{2007}]{schennach2007point}
Schennach, S.~M. (2007).
\newblock Point estimation with exponentially tilted empirical likelihood.
\newblock {\em The Annals of Statistics\/}~{\em 35\/}(2), 634--672.

\bibitem[\protect\citeauthoryear{Suss and Oliveira}{Suss and
  Oliveira}{2023}]{suss2023}
Suss, J. and T.~Oliveira (2023).
\newblock Economic inequality and the spatial distribution of stop-and-search
  in london.
\newblock {\em British Journal of Criminology\/}~{\em 63\/}(4), 828--847.

\bibitem[\protect\citeauthoryear{Talts, Betancourt, Simpson, Vehtari, and
  Gelman}{Talts et~al.}{2018}]{talts2018validating}
Talts, S., M.~Betancourt, D.~Simpson, A.~Vehtari, and A.~Gelman (2018).
\newblock Validating {B}ayesian inference algorithms with simulation-based
  calibration.
\newblock {\em arXiv:1804.06788\/}.

\bibitem[\protect\citeauthoryear{Tiratelli, Quinton, and Bradford}{Tiratelli
  et~al.}{2018}]{Tiratelli2018ssdisproportionate}
Tiratelli, M., P.~Quinton, and B.~Bradford (2018).
\newblock {Does stop-and-search deter crime? Evidence from ten years of
  London-wide data}.
\newblock {\em The British Journal of Criminology\/}~{\em 58\/}(5), 1212--1231.

\bibitem[\protect\citeauthoryear{Yiu, Goudie, and Tom}{Yiu
  et~al.}{2020}]{yiu2020inference}
Yiu, A., R.~J.~B. Goudie, and B.~D.~M. Tom (2020).
\newblock Inference under unequal probability sampling with the {B}ayesian
  exponentially tilted empirical likelihood.
\newblock {\em Biometrika\/}~{\em 107\/}(4), 857--873.

\bibitem[\protect\citeauthoryear{Zhao, Keele, Small, and Joffe}{Zhao
  et~al.}{2022}]{zhao2022note}
Zhao, Q., L.~J. Keele, D.~S. Small, and M.~M. Joffe (2022).
\newblock A note on posttreatment selection in studying racial discrimination
  in policing.
\newblock {\em American Political Science Review\/}~{\em 116\/}(1), 337--350.

\end{thebibliography}

\pagebreak
\appendix
	\section{Additional details regarding the Stop \& Search dataset}\label{appn} 
	\subsection{Demographic data on ethnic groups}
	
	The demographic data are taken from Census 2021 data by the Office for National Statistics (ONS) and are available from \url{https://www.nomisweb.co.uk/sources/census_2021}. The demographic data features have been collocated at local authority, ``borough", level as the analytical geographical unit in this research.
	
	The ethnic groups in Census 2021 are mainly 5 groups including ``Asian, Asian British, Asian Welsh", ``Black, Black British, Black Welsh, Caribbean or African", ``Mixed or Multiple", ``White", and "Other ethnic group" (ONS, 2021).  To be consistent with the categories of ethnic groups in the Stop \& Search dataset, we further group the ethnic groups into four groups as: ``Asian, Asian British, Asian Welsh", ``Black, Black British, Black Welsh, Caribbean or African", ``White", and ``Others" (including Mixed and Other groups in the official Census dataset). From the latest Census 2021 data, ethnic groups' proportions in London are unevenly distributed in that, White people take up 53.73\%, followed by Asian people at 20.7\%, Black people at 13.52\%, and Others at 12.05\%.
	
	Exploratory data analysis on Stop \& Search subjects' demographics, especially the ethnic compositions, has found disproportionally distributed stop and search events among the ethnic groups. White people took up on average 40\% over time (ranged from 38\% to 42\%), Black people took up on average 38\% during the observation period (ranged from 36\% to 40\%), Asian people took up on average 17\% (ranged from 14\% to 18\%), then Others at 5\% (ranged from 4\% to 6\%).
	
	\begin{figure}[ht]
		\centering
		\includegraphics[width=0.7\textwidth]{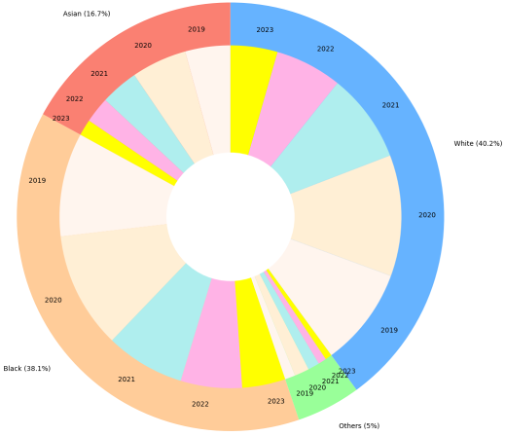}
		\caption{Stop \& Search Subjects Ethnic Groups Proportions}
	\end{figure}
	
	\subsection{Data aggregation procedures}
	
	The data have been sourced from the Metropolitan Police and City of London police force and published in the Single Online Home National Digital Team under Open Government Licence v3.0. They are publicly available from \url{https://data.police.uk/docs/method/stops-at-location/}. Before publishing, the location coordinates of for each stop are anonymized (detailed methods of anonymization from the publisher can be found at \url{https://data.police.uk/about/#location-anonymisation}) and the age of the person stopped has been adjusted to a corresponding age group (e.g. 18-24). 
	
	In this research, 2.6 million stop and search incidents have been compiled for the observation period from January 2019 to December 2023, for seven legislative grounds (three of which are listed in Table \ref{tab:legislative}) for 11 types of Searched Objects. Upon the aggregation of the data into three listed categories, drug objects related stop and search incidents took up 64\% over the observation period, as compared to 12\% for acquisitive objects and 24\% for expressive objects. The searched subjects' ethnic information have been collected from ``officer defined ethnic group".
	
	\begin{table}[ht]
		\caption{Legislative bases underlying the aggregation of stop and search interactions}
		\label{tab:legislative}
		\begin{center}
			\small
			\begin{tabular}{ | m{9em} | m{12em}| m{18em} | } 
				\hline
				\textbf{Category} & \textbf{Searched Objects} & \textbf{Legislation} \\
				\hline
				Acquisitive objects & `Stolen goods', `Article for use in theft' & `Police and Criminal Evidence Act 1984 (section 1)', `Criminal Justice Act 1988 (section 139B)', `Police and Criminal Evidence Act 1984 (section 6)'\\ 
				\hline
				Expressive objects & `Offensive weapons', `Anything to threaten or harm anyone', `Firearms', `Fireworks', `Evidence of offences under the Act', `Crossbows', `Game or poaching equipment' & `Police and Criminal Evidence Act 1984 (section 1)', 'Criminal Justice Act 1988 (section 139B)', `Police and Criminal Evidence Act 1984 (section 6)', `Criminal Justice and Public Order Act 1994 (section 60)', `Firearms Act 1968 (section 47)' \\ 
				\hline
				Drug objects & `Controlled drugs', `Psychoactive substances' & `Misuse of Drugs Act 1971 (section 23)', `Psychoactive Substances Act 2016 (s36(2))' \\ 
				\hline
			\end{tabular}
		\end{center}
	\end{table}

\end{document}